\documentclass[aps,pra,twocolumn,amsmath,showpacs]{revtex4}
\usepackage[T1]{fontenc}        
\usepackage{ae}                 
\usepackage{graphicx}           

\graphicspath{{images/}}         

\begin{document}
\title{Electronic Noise in Optical Homodyne Tomography}

\author{J\"urgen Appel, Dallas Hoffman, Eden Figueroa, A. I. Lvovsky}

\affiliation{Institute for Quantum Information Science, University of
  Calgary, Calgary, Alberta T2N 1N4,
  Canada\footnote{\url{http://www.iqis.org/}}}

\date{\today}

\begin{abstract}
The effect of the detector electronic noise in an optical homodyne tomography experiment is shown
to be equivalent to an optical loss if the detector is calibrated by measuring the quadrature noise
of the vacuum state. An explicit relation between the electronic noise level and the equivalent
optical efficiency is obtained and confirmed in an experiment with a narrowband squeezed vacuum
source operating at an atomic rubidium wavelength.
\end{abstract}

\pacs{42.50.Dv,03.65.Wj,05.40.Ca}

\maketitle

\paragraph{Introduction} \label{sec:Introduction}
Optical homodyne tomography (OHT) is a method of characterizing quantum states of light by measuring the quantum noise
statistics of the electromagnetic field quadratures at different phases. Theoretically
proposed in \cite{VogelRisken} and first experimentally implemented in the early 1990s \cite{smi93a},
OHT has become a standard tool of quantum technology of light, in particular in quantum information
applications.

OHT is prone to a variety of inefficiencies, of which the main one --- the optical losses (OLs) ---
is due to absorption in beam paths, non-unitary efficiency of the photodiodes, imperfect balancing
of the detector, and, partially, imperfect mode matching between the signal and the local
oscillator modes \cite{ray95}. Also, the presence of non-mode-matched signal light appears as added
noise in the detector, raising the apparent noise floor \cite{ray95}. All OLs can be modeled by an
absorber in the signal beam path. The effect of OL in quantum state reconstruction is well
understood, and there are ways of both its quantitative evaluation and compensation for its effects
in the quantum state reconstruction procedure \cite{leonhardt,kiss95}.

An important additional inefficiency source is the electronic noise (EN) of the homodyne detector.
Its effect is the addition of a random value to each quadrature measurement, which blurs the marginal
distribution and causes errors in quantum state reconstruction. Unlike the optical inefficiency,
the EN, to our knowledge, has not yet been studied, and is usually neglected whenever an efficiency
analysis is made
\cite{smi92,Fock,DispFock,QubitHT,our06,Fock2,nee06}.

The purpose of this paper is to fill this gap and to analyze the role of the electronic noise in
OHT. We show its effect to be equivalent to that of optical losses. The key to this equivalence is
the detector calibration procedure, which involves measuring the quadrature noise of the vacuum
state. Applying this procedure to the vacuum state contaminated with the EN results in rescaling
of the field quadratures, analogous to that taking place due to optical absorption.

We confirm our experimental findings in a squeezing experiment, in which we vary the local
oscillator power to achieve different signal-to-noise ratios of the homodyne detector. The squeezed
vacuum source is a narrowband, tunable continuous-wave optical parametric amplifier operating at a
795-nm wavelength.

We begin our theory by writing the effects of both inefficiency types on the reconstructed Wigner
function (WF).

\paragraph{The beam splitter model for optical signal loss} \label{sec:Part I}
According to the beam splitter model of absorption, non-unitary optical efficiency $\eta$ modifies
the WF $W(X,P)$ of the signal state as follows \cite{leonhardt}:
\begin{eqnarray} \label{opt}
 W_{OL}(X,P)&=&\frac{1}{\pi\eta(1-\eta)}\iint   W(\eta^{-1/2}
X^\prime,\eta^{-1/2} P^\prime)
\\ \nonumber&\times& \exp{\left[-\frac{(X-X^\prime)^2+(P-P^\prime)^2}{1-\eta}\right]}dX^\prime dP^\prime.
\end{eqnarray}
This transformation can be understood by visualizing the
absorber as a fictional beam splitter with the signal beam entering one port and the vacuum state
entering the other. Transmission of the signal field through the beam splitter rescales the field
amplitudes by a factor of $\eta^{-1/2}$. The vacuum state reflected from the beam splitter into the signal mode adds noise, which expresses as a convolution of the WF with a
two-dimensional Gaussian function of width $\sqrt{1-\eta}$.

\paragraph{Effect of the electronic noise} \label{sec: Part II}
In the absence of the electronic noise, the electric signal (voltage or current) $V$ generated by the homodyne detector is proportional to the quadrature sample $X$: $V=\alpha X$. The coefficient $\alpha$ has to be determined when quantum state reconstruction is the goal. This is usually done by means of a calibration procedure, which consists of acquiring a set of quadrature noise samples $\{V_{0i}\}$ of the vacuum state, prepared by letting only the local oscillator into the detector. The vacuum marginal distribution,
\begin{equation}
\label{pr0}{\rm pr}_{0}(X)=\pi^{-1/2}\exp{(-X^{2})},
\end{equation} is loss-independent, and has a mean square of $\langle X^2\rangle=1/2$. One can thus use the experimental data to obtain
\begin{equation}\label{meanV0}
\alpha=\sqrt{2\langle V_{0i}^2\rangle}.
\end{equation}

When the EN is present, the observed probability distribution of the electric signal is a convolution of the scaled marginal distribution ${\rm pr}(X)$ of the state measured and the electronic noise histogram  $p_e(V)$:
\begin{equation}\label{prEN}
\tilde{\rm pr}_{EN}(V)=\frac{1}{\alpha}\int{\rm pr}\left(\frac{V'}{\alpha}\right)\, p_e (V-V') \,dV'
\end{equation}
(the factor of $\alpha$ is added to retain the normalization of the marginal distribution).
 Under realistic experimental conditions of a broadband electronic noise, its distribution can be assumed Gaussian:
\begin{equation}\label{pe}
p_e(V)=\frac{1}{\sqrt{\pi T}}\exp\left(-\frac{V^2}{T}\right),
\end{equation}
where $T$ is the mean square noise magnitude.

Consider the detector calibration procedure in the presence of the EN. According to Eqs. (\ref{pr0}), (\ref{prEN}), and (\ref{pe}), the vacuum state measurement yields a distribution
\begin{equation}\label{prENvac}
\tilde{\rm pr}_{0,EN}(V_0)=\frac{1}{\sqrt\pi\sqrt{\alpha^2+T}}\exp\left(-\frac{V_0^2}{\alpha^2+T}\right)
\end{equation}
with a mean square of $\langle V_0^2\rangle=(\alpha^2+T)/2$. If no correction for the EN is made, the quantity
\begin{equation}
\alpha'=\sqrt{2\langle V_0^2\rangle}=\sqrt{\alpha^2+T}
\end{equation}
is interpreted according to Eq.~(\ref{meanV0}) as the conversion factor between the electric signal and the field quadrature.

In a homodyne measurement of an unknown state, one acquires a set of samples from the homodyne detector, and then applies the calibration factor $\alpha'$ to rescale its histogram $\tilde{\rm pr}_{EN}(V)$ and obtain the associated marginal distribution. Using Eq.~(\ref{prEN}), we write
\begin{eqnarray}\label{prENq}
{\rm pr}_{EN}(X)&=&\alpha'\tilde{\rm pr}_{EN}(\alpha'X)\\ \nonumber
&=&\frac{\alpha'}{\alpha}\int{\rm pr}\left(\frac{V'}{\alpha}\right)\, p_e (\alpha'X-V') \,dV'.
\end{eqnarray}
Introducing a new integration variable $X'=V'/\alpha'$ and utilizing the EN distribution (\ref{pe}), Eq.~(\ref{prENq}) can be rewritten as
\begin{eqnarray}
{\rm pr}_{EN}(X)&=&\frac{\alpha'^2}{\sqrt{\pi T}\alpha}\int{\rm pr}\left(\frac{\alpha'}{\alpha}X'\right)\,\\ \nonumber &\times& \exp\left[-\frac{(X-X')^2}{T/\alpha'^2}\right] \,dX'.
\end{eqnarray}
This rescaled, noise contaminated marginal distribution is then used to reconstruct the Wigner function.

Because the marginal distribution is tomographically linked to
the WF,
\begin{equation}
{\rm pr}_\theta(X)=\int W(X\cos{\theta}-P\sin{\theta},X\sin{\theta}+P\cos{\theta})\,dP,
\end{equation}
the Wigner function reconstructed from a noisy measurement is a convolution
\begin{eqnarray} \label{elec}
\nonumber W_{EN}(X,P)&=&\frac{\alpha'^4}{\pi T \alpha^2}\iint dX^\prime dP^\prime \ W\left(\frac{\alpha'}{\alpha}X',\frac{\alpha'}{\alpha}P'\right)
\\ &\times& \exp{\left[-\frac{(X-X^\prime)^2+(P-P^\prime)^2}{T/\alpha'^2}\right]}.
\end{eqnarray}

\paragraph{Equivalence of the efficiency loss mechanisms} \label{sec: Part III}
Comparing the above expression with Eq. (\ref{opt}) for the WF of a state affected by optical losses, we find that the two are equivalent if we set
\begin{equation} \label{result}
\eta_{eq}=\frac{\alpha^2}{\alpha'^2}=\frac{\alpha^2}{\alpha^2+T}.
\end{equation}
In other words, we may treat the electronic noise as an additional optical attenuator with a transmission equal to $\eta_{eq}$. Accordingly, the known numerical procedures for correcting for optical losses can be readily applied to the electronic noise.

This result may appear surprising. The field quadrature noise measured in the presence of the EN is a convolution of the WF's marginal distribution with the noise histogram. In the case of OL, on the other hand, the convolution is preceded by rescaling of the quadrature variables. The similarity between the effects of the two inefficiency emerges due to the detector calibration procedure. The vacuum state measurement is distorted by the EN but not the OL, hence there is additional rescaling of the quadratures in the presence of the EN.

For experimental applications, it is more convenient to express the equivalent efficiency in terms of the signal-to-noise ratio of the detector. By the latter we understand the ratio between the observed mean square noise of the vacuum state (in the presence of the EN) and the mean square electronic noise, i.e. the quantity
\begin{equation} \label{ST}
S=\frac{\alpha^2+T}{T}.
\end{equation}
Re-expressing Eq.~(\ref{result}) in terms of $S$ yields
\begin{equation} \label{resultS}
\eta_{eq}=\frac{S-1}{S}.
\end{equation}
As evidenced by this equation, reasonably low electronic noise does not constitute a significant
efficiency loss. For example, the original time-domain detector of Smithey and co-workers
\cite{smi92,smi93a}, featuring a 6 dB signal-to-noise ratio (i.e. the rms quadrature noise being
only twice as high as the rms EN) has an equivalent quantum efficiency of 75\%. A similar
equivalent efficiency is featured by a high-speed detector by Zavatta {\it et al.}
\cite{ZavattaHD}. A 14-dB time-domain detector of Hansen and co-workers \cite{HansenHD} has an
equivalent efficiency of 96\%, and a typical frequency-domain detector with a signal-to-noise ratio
of 20 dB has an efficiency of 99\%.

\paragraph{Experiment}
We verify our theoretical findings in a squeezing experiment. The squeezed vacuum source is a
continuous-wave optical parametric amplifier (OPA), operating at the 795-nm wavelength of the D1
transition in atomic rubidium. We have constructed this source to perform experiments on
interfacing quantum information between light and atoms, in particular storage
\cite{kuzmich:05,lukin:05} and Raman adiabatic transferring \cite{appel:06,vewinger-2006} of light
in atomic vapor. Similar sources have been reported very recently by two groups
\cite{tanimura-2006,hetet-2006}

The optical cavity of the OPA is of bow-tie configuration composed of four mirrors, among which two
are flat and two are concave with a curvature radius of 100 mm. The cavity is singly resonant for the
generated 795-nm wavelength. As the output coupler we use one of the flat cavity mirrors, which had
a reflectivity of 93\%. Other cavity mirrors had an over 99.9\% reflectivity. With these
parameters, the cavity had a 460-MHz three free spectral range with the resonant line width of 6 MHz

The source employs a 5-mm long periodically poled KTiOPO$_4$ crystal as the nonlinear element. The
crystal, manufactured by Raicol, features a 3.175 $\mu$m poling period and double antireflection
coating on both sides. It is placed at the beam waist between the curved mirrors.

The OPA is pumped by up to 240 mW of 397.5 nm light. The pump field is the second harmonic of a frequency stabilized titanium-sapphire
master laser. The same laser is used to provide the local oscillator for the homodyne detector. The OPA cavity is locked at resonance by means of an additional diode laser phase locked to the master oscillator at 3 free spectral ranges of the OPA cavity. The cavity lock is implemented using the Pound-Drever-Hall method,
using a beam modulated at 20 MHz and propagating in the direction opposite to the pumped mode.

\begin{figure}
  \centering
  \includegraphics[width=0.9\columnwidth,keepaspectratio]{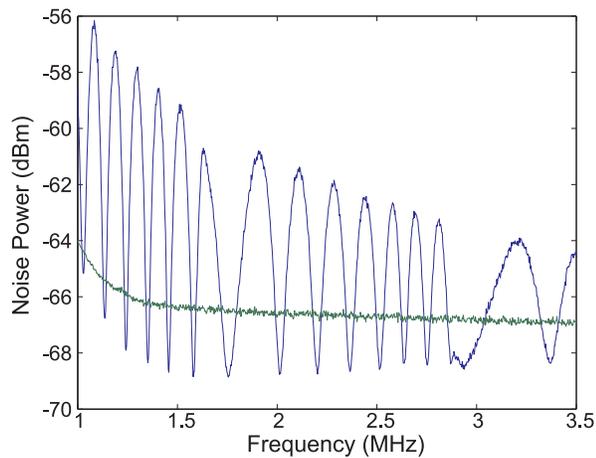}
  \caption{Quadrature noise spectrum of the squeezed vacuum generated by the OPA, obtained by a simultaneous scan of the local oscillator phase and the detection frequency. The green line is the shot noise spectrum.}
\end{figure}

The circuit for homodyne detection uses two Hamamatsu S3883 photodiodes of 94\% quantum efficiency
in a ``back-to-back" configuration, akin to that employed in Refs. \cite{HansenHD,ZavattaHD}, followed by a OPA847 operational preamplifier. The detector exhibits
excellent linearity and an up to 12-17 dB signal-to-noise ratio over a 100-MHz bandwidth, with local
oscillator powers up to 10 mW. At a 1-MHz sideband relevant for this experiment, the highest
signal-to-noise ratio of the detector exceeds 17 dB, which, according to Eq.~(\ref{resultS}),
corresponds to equivalent optical losses of only 2\%. The detector output is observed in the frequency
domain using an Agilent ESA spectrum analyzer.

Figure 1 shows the electronic spectrum of squeezed vacuum field generated by the OPA. Up to 3 dB of
squeezing is observed within the cavity linewidth.

To verify the predictions of Eq.~(\ref{resultS}), we stabilize the pump level at 140 mW and set the
spectrum analyzer to run in the zero span mode at a detection frequency of 1 MHz. The signal-to-noise ratio of the
detector is varied by changing the local oscillator intensity. At each intensity value, we record
three spectrum analyzer traces: electronic noise (measured by blocking both photodiodes of the
homodyne detector), shot noise (measured by blocking the squeezed vacuum signal) and the
phase-dependent squeezed vacuum quadrature. The highest ($\langle Q_+^2\rangle$) and lowest
($\langle Q_-^2\rangle$) quadrature noise levels of the squeezing measurement are then determined
and normalized to the shot noise level. The uncertainty of $\langle Q_\pm^2\rangle$ is estimated as 0.2
dB; it is much smaller for the shot-noise and electronic-noise data. The results of these
measurements are shown in Fig.~2.

\begin{figure}
  \centering
  \includegraphics[width=0.9\columnwidth,keepaspectratio]{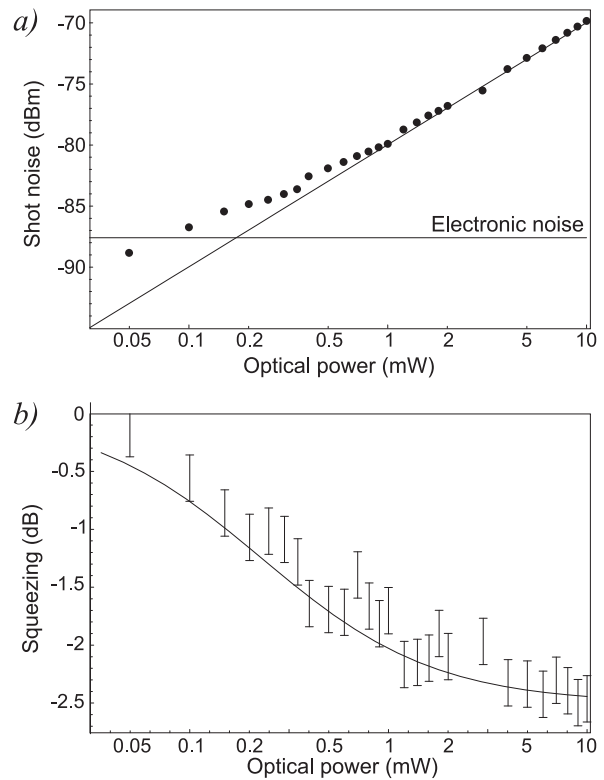}
  \caption{Experimental results. a) Observed shot noise as a function of the local oscillator power. Also
  displayed is a linear fit and the electronic noise level. b) Observed squeezing as a function of the local oscillator power. The solid line is calculated using Eq.~(\ref{prEN}).}
\end{figure}

To determine the quantum efficiency, we use the fact that for the pure squeezed state, the product
of the mean square quadrature uncertainties is the minimum allowed by the uncertainty principle,
i.e. $\langle Q_{{\rm pure},+}^2\rangle\langle Q_{{\rm pure},-}^2\rangle=1/4$, where the
normalization factor corresponds to the shot noise level. After undergoing an optical loss, the
squeezed state remains Gaussian, but loses its purity. Using the beam splitter model of absorption,
we find that upon propagating through an attenuator of transmission $\eta$, the quadrature noise
reduces by a factor of $\eta$, but gains excess vacuum noise of intensity $1-\eta$:
\begin{subequations}\label{Qsq}
\begin{eqnarray}
\langle Q_+^2\rangle=\langle\eta Q_{{\rm pure},+}^2\rangle+ (1-\eta)/2; 
\\ \langle Q_-^2\rangle=\langle\eta Q_{{\rm pure},-}^2\rangle+ (1-\eta)/2.
\end{eqnarray}
\end{subequations}
Measuring $\langle Q_+^2\rangle$ and $\langle Q_-^2\rangle$ and solving Eqs. (\ref{Qsq}), we can
find the value of $\eta$, i.e. how much loss a pure squeezed state would have experienced to
generate the state with the quadrature noise levels observed \cite{lwb06b}:
\begin{equation}
\eta=\frac{(2\langle Q_+^2\rangle-1)(1-2\langle Q_-^2\rangle)}{2\langle Q_+^2\rangle+2\langle
Q_-^2\rangle-2}.
\end{equation}
 The result of applying this method to
our squeezed vacuum data is displayed in Fig.~3 along with a fit with Eq.~(\ref{resultS})
multiplied by an overall efficiency factor of 0.51 (emerging due to various optical losses). The
experimental data matches well to the theoretical prediction.

\paragraph{Conclusion} \label{conclusion}
We have found that there is a strong similarity between the effects of optical absorption and
electronic noise on the state reconstructed by means of optical homodyne tomography. Therefore, the
detector's Gaussian electronic noise can be treated as an additional optical loss which is related
to the detector's signal-to-noise ratio by Eq. (\ref{resultS}). The finding is
verified in a squeezing experiment involving a narrowband nonclassical light source suitable for
experiments on interfacing quantum information between light and atoms.

\paragraph*{Acknowledgements}
We thank Michael Raymer for discussions. A. L.'s research is supported by Quantum\emph{Works}, NSERC, CFI, AIF, and CIAR.

\begin{figure}[t]
  \centering
  \includegraphics[width=0.9\columnwidth,keepaspectratio]{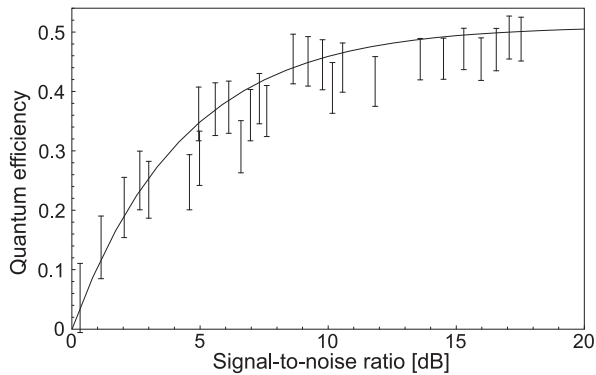}
  \caption{Equivalent quantum efficiency $\eta_{eq}$ as a function of the detector's signal-to-noise ratio $S$.
  This ratio is expressed in decibels as $10\log_{10}S$. Both the experimental results and the
  theoretical preediction (\ref{resultS}) (multiplied by an optical efficiency factor of 0.51) are
  displayed.}
\end{figure}

\bibliography{error8}

\end{document}